\documentclass{article}

\usepackage{PRIMEarxiv}

\usepackage[utf8]{inputenc} 
\usepackage[T1]{fontenc}    
\usepackage{hyperref}       
\usepackage{url}            
\usepackage{booktabs}       
\usepackage{amsfonts}       
\usepackage{nicefrac}       
\usepackage{microtype}      
\usepackage{lipsum}
\usepackage{fancyhdr}       
\usepackage{graphicx}       
\usepackage{tabularx}
\usepackage{threeparttable} 
\usepackage{makecell}
\usepackage{placeins}
\usepackage[table]{xcolor}
\graphicspath{{media/}}     

\title{Topological Determinants of Resilience in Urban Rail Networks Facing Multi-Hazard Disruptions
}

\author{
  Ashis Kumar Pal \\
  Northeastern University \\
  Boston, MA, 02115\\
  \texttt{pal.as@northeastern.edu} \\
   \And
  Auroop R. Ganguly \\
  Professor, CEE \\
  Northeastern University, \\
  Boston, MA, 02115\\
  \texttt{a.ganguly@northeastern.edu} \\
}

\begin{document}
\maketitle

\begin{abstract}
This study examines the failure and recovery, two key components of resilience of nine major urban rail networks - Washington DC, Boston, Chicago, Delhi, Tokyo, Paris, Shanghai, London, and New York - against multi-hazard scenarios utilizing a quantitative approach focused on topological parameters to evaluate network resilience. Employing percolation-based network dismantling approach like Sequential Removal of Nodes and Giant Connected Component analysis, alongside random, centrality-based targeted attacks and flooding failure, findings reveal Domirank centrality's superior resilience in disruption and recovery phases. Kendall's tau coefficient's application further elucidates the relationships between network properties and resilience, underscoring larger networks' vulnerability yet faster recovery due to inherent redundancy and connectivity. Key attributes like average degree and path length consistently influence recovery effectiveness, while the clustering coefficient's positive correlation with recovery highlights the benefits of local interconnectivity. This analysis emphasizes the critical role of select nodes and the importance of balancing network design for enhanced resilience, offering insights for future urban rail system planning against multi-hazard threats.
\end{abstract}

\keywords{Urban Rail Network Resilience \and Topological attributes \and Multi-Hazard Threats \and Strategic Node Importance in Rail Networks.}

\section{Introduction}
The interconnected Urban rail transportation networks are intricate in design and critical for metropolitan functionality which serves as the circulatory system of metropolitan areas, enabling the flow of millions daily. The structural integrity and operational efficiency of these networks are not merely matters of convenience but are essential for urban resilience and sustainability. They are influenced by an array of topological metrics,  such as average degree, average shortest path length, network density, diameter, modularity index, and clustering coefficient commonly used to describe a network's architecture as well as their functional robustness and vulnerability to disruptions (\cite{boccaletti2006complex}).

Complex networks theory, which has its roots in the work of (\cite{erdHos1961strength}), has revolutionized our understanding of various interconnected systems. The advent of this theory has brought about a nuanced appreciation of network behaviors, elucidating the intricate relationship between structure and function. The subsequent research by Watts and Strogatz (\cite{watts1998collective}) and Barab\'asi and Albert (\cite{barabasi1999emergence}) has extended this foundation, exploring the dynamic nature of real-world networks, including urban rail systems. These networks often display small-world properties, characterized by a dense web of local connections and surprisingly short paths across the entire network.

The small-world phenomenon, first detailed in the networks of the Boston subway system(MBTA)  by Latora and Marchiori (\cite{latora2002boston}) and further confirmed in city street networks by Jiang and Claramunt (\cite{jiang2004topological}), illustrates a potent combination of local and global efficiency in network design. This dual efficiency facilitates quick and versatile responses to everyday demands and extraordinary stresses alike. Likewise, an examination of the worldwide airport network revealed similar small-world dynamics, where high clustering coefficients suggest an innate tendency for node groupings, and low average shortest path lengths ensure operational efficiency across vast geographical expanses (\cite{amaral2000classes}; \cite{guimera2005worldwide}). The work by (\cite{xu2023interconnectedness} expands knowledge by systematically comparing the resilience of five distinct metro networks. Their analysis suggests that the superior resilience of the Singapore metro may be attributed to its moderately heterogeneous topology and efficient flow patterns, which contribute to the smallest average travel distance observed among the networks studied.

The average shortest path length and clustering coefficient, specifically within urban rail networks, have a direct bearing on the passenger experience, dictating the ease of movement and the robustness of the service network. These parameters signify more than mathematical abstractions; they represent the tangible connections between communities and the very accessibility of urban opportunities. In a well-connected network, these topological features foster resilience by enabling alternative pathways for movement even in the event of partial network failures.

While numerous studies have explored the topological characteristics of transportation networks, a parallel line of inquiry has emerged, focusing on the resilience of these networks amidst various types of perturbations. The Giant Connected Component (GCC) has emerged as a vital measure of network functionality, particularly in assessing network integrity under disruptive events. The work by \cite{xu2021resilient} expands our knowledge by systematically comparing the resilience of five distinct metro networks. Their analysis suggests that the superior resilience of the Singapore metro may be attributed to its moderately heterogeneous topology and efficient flow patterns, which contribute to the smallest average travel distance observed among the networks studied. It is here, in the nexus between static topological measures and dynamic resilience, that an intriguing gap exists in current literature. The relationship between these enduring network characteristics and the network's ability to withstand and recover from disruptions—reflected quantitatively in the area under the failure and recovery curves—remains underexplored.

Numerous studies in this domain have highlighted the susceptibility of urban rail networks to targeted disruptions. For instance, Daniel Sun and Yuhan Zhao's investigation of Shanghai's Urban rail transit networks suggests that networks are particularly vulnerable to attacks on nodes of high degree and betweenness—a finding that may not always correlate with passenger volume (\cite{yuan2020passenger}). \cite{li2019resilience} further explored this with their analysis of the Beijing Subway, reinforcing the notion of selective vulnerability within these complex systems. The work of \cite{piraveenan2012assortativity} extends the discourse to the robustness of networks, especially under sustained targeted attacks, revealing the relative fragility of scale-free networks as compared to their robustness against random failures.

Building on these foundations, this study aims to contribute a substantive expansion to the field by simulating disruptions and recovery across a diverse array of urban rail networks, each with distinct sizes and topological features. We perform a comparative analysis using DomiRank centrality introduced by \cite{engsig2024domirank}, an state of the art measure for evaluating node importance, benchmarking it against conventional centrality measures to assess its effectiveness in predicting network resilience.

As we seek to unravel the complex interplay between network topology and resilience, where the size of a network, its density, average degree, average shortest path length, and its clustering coefficient, along with the modularity index, are static descriptors of a network's structure. This research departs from the typical comparative analyses with random graphs that traditionally underscore small-world features in transportation networks. It embarks on a distinct trajectory—measuring resilience and recovery patterns through the prism of topological impacts. In exploring these dynamics across nine diverse urban rail networks spanning from Washington DC, Boston, Chicago, Delhi, Tokyo, Paris, Shanghai, London, to New York, our study employs a quantitative approach. It evaluates the effects of five types of centrality-based ( Degree centrality, Betweenness centrality, Closeness centrality, Eigenvector centrality, DomiRank centrality) attacks and recoveries. By simulating both random failures and flooding, alongside targeted disruptions based on centrality measures, we methodically prioritize stations for sequential disruption and recovery, illuminating the nuanced ways in which network metrics can influence failure and recovery processes. In doing so, we aspire to shed light on how network metrics influence the dynamics of failure and recovery within urban rail systems and, ultimately, inform the development of resilient transportation infrastructures for the future.

\begin{figure}[!ht]
\centering
\includegraphics[width=1\linewidth]{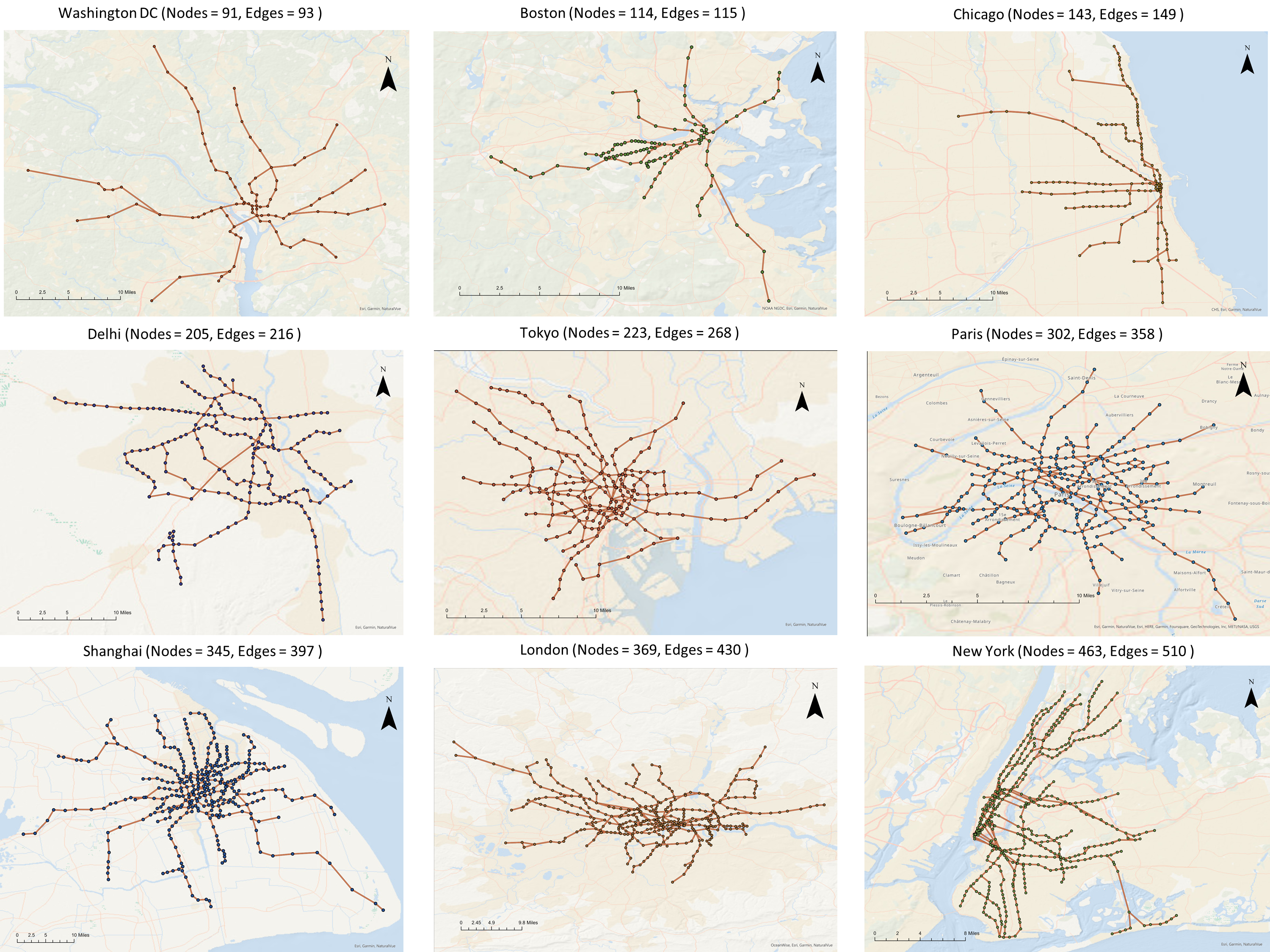}
\caption{\label{fig:Urban}This figure presents a comparative visualization of nine urban rail networks, arranged in ascending order of nodes(N) and edges(E). The networks span across diverse global regions, including Washington DC, Boston, Chicago, Delhi, Tokyo, Paris, Shanghai, London, and New York. Each network is depicted within a distinct subplot, showcasing the intricate layout of rail lines and stations within each city. Light blue water bodies highlight the presence of rivers, coasts, or lakes, providing context to the geographical setting of these urban transportation systems.
}
\end{figure}

Within the framework of urban rail systems, characterized by the Urban Rail Infrastructure Systems (URIS), several key topological attributes emerge as instrumental in understanding their structure and potential performance. The nodes, represented by \(N\), define the total number of stations in each system, indicating its scale and complexity. Complementing the nodes, the edges, \(E\), represent the direct routes between stations, forming the backbone of connectivity across the network. The average degree, denoted by \(\langle k \rangle\), reflects the average number of connections per node, serving as an indicator of network redundancy and accessibility. 

The efficiency and navigational ease within the network are encapsulated by the average shortest path length, \(\langle L \rangle\), measuring the mean number of stops required for travel between station pairs. Network density, symbolized by \(D\), provides insight into the compactness of the network, quantifying the proportion of potential connections that are actualized. The network diameter, \(\Phi\), marks the longest distance between two stations in the shortest path terms, offering a perspective on the spatial extent of the network.

Furthermore, the modularity index, \(M\), gauges the presence of modular configurations, which can hint at the system's subdivision into distinct, internally dense clusters. Lastly, the clustering coefficient, \(C\), reveals the propensity for stations to cluster together, a feature that often facilitates robust local service continuity in the face of disruptions. Each of these attributes plays a pivotal role in the resilience and robustness analysis of urban rail networks, setting the stage for a comprehensive study of infrastructure sustainability.

\begin{table}[!htbp]
\centering
\caption{URTS Systems Characteristics}
\begin{threeparttable}
\begin{tabularx}{\textwidth}{@{}lXXXXXXXl@{}}
\toprule
URTS & Nodes & Edges & \makecell{Avg.\\ Deg.} & \makecell{Avg.\\ SPL} & \makecell{Net.\\ Den} & \makecell{Net.\\ Dia} & \makecell{Mod.\\Ind} & \makecell{Clus.\\Coeff} \\
\midrule
Washington DC & 91  & 93   & 2.04 & 11.51 & 0.0227 & 28.00 & 0.76 & 0.0000 \\
Boston        & 114 & 115  & 2.02 & 13.95 & 0.0179 & 34.00 & 0.81 & 0.0058 \\
Chicago       & 143 & 149  & 2.08 & 15.99 & 0.0147 & 43.00 & 0.82 & 0.0212 \\
Delhi         & 205 & 216  & 2.11 & 17.08 & 0.0103 & 54.00 & 0.82 & 0.0000 \\
Tokyo         & 223 & 268  & 2.40 & 10.64 & 0.0108 & 32.00 & 0.77 & 0.0233 \\
Paris         & 302 & 358  & 2.37 & 11.98 & 0.0079 & 34.00 & 0.80 & 0.0157 \\
Shanghai      & 345 & 397  & 2.30 & 16.00 & 0.0067 & 45.00 & 0.83 & 0.0086 \\
London        & 369 & 430  & 2.33 & 13.73 & 0.0063 & 39.00 & 0.83 & 0.0294 \\
New York      & 463 & 510  & 2.20 & 19.69 & 0.0048 & 57.00 & 0.88 & 0.0279 \\
\bottomrule
\end{tabularx}
\begin{tablenotes}
\item Table 1: Topological Characteristics of Urban Rail Transit Systems (URTS). This table details the structural properties of the nine urban rail transit systems, comparing key topological metrics such as number of nodes and edges, average degree, average shortest path length (Avg. SPL), network density, network diameter, modularity index, and clustering coefficient. The data illustrates the complexity and connectivity within each urban rail network, providing a foundational comparison for further analysis of network robustness and resilience.

\end{tablenotes}
\end{threeparttable}
\end{table}

\section{Methodology:}\label{sec2}

Simulating network performance under diverse disruptive events, such as natural disasters, operational accidents, and terrorist attacks, often involves the application of random failures and targeted attacks [cite]. In the context of this study, random failure refers to the random disruption of stations, one by one, with an equal probability of failure assigned to each station. Targeted attacks involve a strategic sequence where stations are systematically attacked based on a predefined strategy. However, executing such attacks necessitates knowledge of the entire network structure. Centrality in the network is fundamentally defined as the measure of a node's importance relative to other nodes. This study utilizes five types of centrality-based attacks, which include newer domirank-centrality-based attacks, closeness-based attacks, betweenness-based attacks, degree-based attacks, and eigenvector-based attacks, to simulate targeted attacks. In the context of centrality-based attacks, stations are disrupted sequentially from high to low, prioritized according to their centrality values.

\subsection{Random Failure:}
Random failure is simulated by probabilistically removing stations from the network without any specific pattern or preference, mimicking the effects of unpredictable events that cause station closures. This approach assumes a uniform likelihood of failure across all stations, thereby providing a baseline scenario to understand the inherent resilience of the network. For the simulation, 20 independent trials are conducted where stations are eliminated randomly, and the ensemble mean of the Giant Connected Component (GCC) is calculated to evaluate the overall connectivity and robustness of the network post-failure.

\subsection{Targeted Attack:}\label{subsec1}

\subsubsection{Domirank Centrality:}\label{subsubsec1}
DomiRank Centrality, when applied to an urban rail transportation network, is a metric that quantifies the relative importance and influence of a particular station within the network based on its dominance over its neighboring stations. It evaluates each station's ability to influence the flow and connectivity of the network, particularly focusing on its dominance in relation to nearby, less significant stations. DomiRank centrality, denoted as \(\Gamma\), is defined as the stationary solution of the dynamical process given by:

\[
\frac{d\Gamma(t)}{dt} = \alpha A(\theta 1_{N \times 1} - \Gamma(t)) - \beta\Gamma(t)
\]

where \(A\) is the adjacency matrix of the network, \(\alpha\) and \(\beta\) are parameters, \(\theta\) is a threshold parameter, and \(1_{N \times 1}\) is an \(N\)-dimensional vector of ones . This dynamical process leads to the analytical expression of DomiRank centrality as:

\[
\Gamma = \theta\sigma(\sigma A + I_{N \times N})^{-1}A1_{N \times 1}
\]

Here, \(\sigma = \frac{\alpha}{\beta}\), \(I_{N \times N}\) is the \(N \times N\) identity matrix, and the convergence interval for \(\sigma\) is defined such that \(\sigma(N) \in \left(0, \frac{1}{- \lambda_N}\right)\), where \(\lambda_N\) represents the minimum (dominant negative) eigenvalue of \(A\). The threshold for domination \(\theta\) plays a role in rescaling the resulting DomiRank centrality, and without loss of generality, can be chosen as \(\theta = 1\).

\subsubsection{Betweenness Centrality:}\label{subsubsec1}
Betweenness Centrality for an urban rail transportation network is a measure that quantifies the importance of a station based on its role in facilitating the shortest paths between pairs of other stations within the network. It is calculated as the ratio of the number of shortest paths passing through a given station to the total number of shortest paths connecting all pairs of stations in the network. Stations with higher Betweenness Centrality values play a crucial role in maintaining efficient connectivity by serving as key points for transit within the urban rail system. For a node \(v\), the betweenness centrality \(C_B(v)\) is defined as:

\[
C_B(v) = \sum_{s \neq v \neq t} \frac{\sigma_{st}(v)}{\sigma_{st}}
\]
where \(\sigma_{st}\) is the total number of shortest paths from node \(s\) to node \(t\) and \(\sigma_{st}(v)\) is the number of those paths that pass through \(v\).

\subsubsection{Degree Centrality:}\label{subsubsec1}
Degree Centrality in an urban rail transportation network quantifies the significance of a station based on the number of direct connections it has with other stations. It is determined by counting the number of immediate links a station possesses, reflecting its accessibility and potential traffic volume. Stations with a high degree centrality are critical hubs within the network, facilitating numerous direct routes and serving as essential points for passenger transfer and distribution throughout the urban rail system. For a node \(v\), the degree centrality \(C_D(v)\) is simply:
\[
C_D(v) = \frac{1}{N - 1} \sum_{u \in V} a_{uv}
\]

Where \( N \) is the total number of nodes in the network, \( V \) is the set of all nodes in the network, \( a_{uv} \) is an element of the adjacency matrix where \( a_{uv} = 1 \) if there is an edge between nodes \( u \) and \( v \), and \( a_{uv} = 0 \) otherwise.

\subsubsection{Closeness Centrality:}\label{subsubsec1}
Closeness Centrality in an urban rail transportation network assesses the relative proximity of a station to all other stations within the network. It is determined by inverting the sum of the shortest path distances from a given station to every other station, reflecting how easily and quickly a station can be reached. Stations with high Closeness Centrality are strategically positioned to minimize travel time for passengers, making them integral to the efficiency of the rail network by enabling faster and more direct routes to a wide array of destinations. For a node \(v\), the closeness centrality \(C_C(v)\) is defined as:

\[
C_C(v) = \frac{N - 1}{\sum_{u \neq v} d(v, u)}
\]
where \(d(v, u)\) is the distance between nodes \(v\) and \(u\), and \(N\) is the total number of nodes.

\subsubsection{Eigenvector Centrality:}\label{subsubsec1}
Eigenvector Centrality in an urban rail transportation network measures a station's influence by considering not only the number of connections it has but also the significance of those connected stations within the network. It is calculated based on the principle that connections to highly connected and important stations contribute more to a station's centrality than connections to less significant ones. Stations with high Eigenvector Centrality are considered influential within the network, indicating that they are well-connected to other stations that themselves are central and vital to the network's functionality. 
For a node \(v\), if \(\mathbf{A}\) is the adjacency matrix of the network and \(\mathbf{x}\) is the eigenvector of \(\mathbf{A}\) corresponding to the largest eigenvalue, then the eigenvector centrality \(C_E(v)\) is the \(v\)th element of \(\mathbf{x}\).

\[
C_E(v) = \frac{1}{\lambda} \sum_{u \in V} a_{uv}C_E(u), \forall v \in V_t, V_f
\]

where \( C_E(v) \) is the eigenvector centrality for station \( v \), \( \lambda \) is the largest eigenvalue of the adjacency matrix \( A \), \( a_{uv} \) represents the edge weight between stations \( u \) and \( v \), and the sum is taken over all stations \( u \) that are connected to station \( v \). The sets \( V_t \) and \( V_f \) represent different sets of nodes, potentially reflecting different types of networks (like travel-time-weighted and flow-weighted networks).

\subsection{Flooding Failure:}
Flooding failure simulation incorporates geographical information to assess the impact of riverine flooding on the urban rail network. Stations are geolocated and ranked based on their proximity to nearby rivers, with the assumption that those closer to water bodies are more susceptible to flooding. The nodes are then sequentially disrupted, starting from the most vulnerable to the least, representing a realistic flooding event. This first-order model allows for the evaluation of network resilience specifically against flooding by considering the spatial distribution of stations in relation to flood-prone areas.

\section{Results:}\label{sec3}

\subsection{Failure}\label{subsec2}

\begin{figure}[!ht]
\centering
\includegraphics[width=1\linewidth]{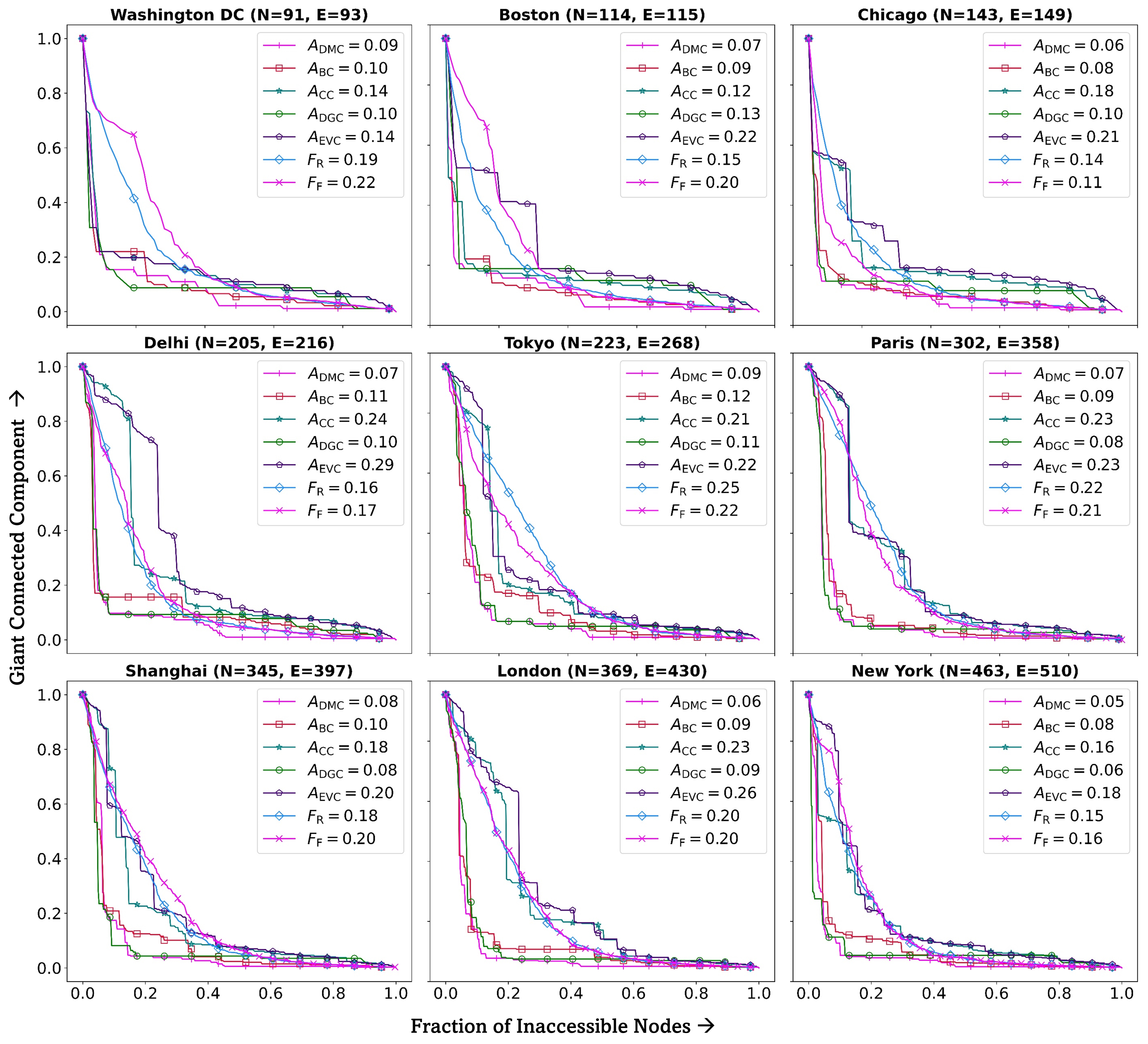}
\caption{\label{fig:Urban}Inter-comparison of targeted attacks and failure curves of the nine urban rail transportation networks (URTS) after perturbations. The y-axis refers to the relative size of the giant-connected component (GCC) as the network loses functionality and x-axis includes the fraction of nodes (stations) failed till that point. A higher AUC reflects higher robustness, and a lower AUC score also indicates higher efficacy of the strategy from an attacker's perspective.}
\end{figure}

we observed that, in general, most networks exhibited their lowest robustness when subjected to a domirank-centrality-based attack. Conversely, they displayed greater resilience when exposed to an eigenvector-centrality-based attack. It's noteworthy, though, that there were specific networks for which a betweenness-centrality and degree-based attack proved to be the most effective in disrupting network functionality.

Flood failure behaves similar to a localized random failure; however, depending on how many ’important’ nodes lie in the flooding zone, the loss of network functionality can be much greater than a baseline random failure. The Chicago L network exhibits the lowest level of resilience to flooding failures among the nine networks studied. Approximately 20-30\% of the nodes crucial for maintaining network connectivity are situated in flood-prone areas. In contrast, the Tokyo metro system displays the highest level of resilience when faced with flooding failures.
Across all nine networks studied, a consistent pattern emerges: a very small portion of nodes (approximately 10\% to 30\%) of the nodes within these networks are pivotal for sustaining connectivity in the face of various types of attacks and failures. This finding underscores the critical role played by a relatively small fraction of nodes in maintaining network integrity under adverse conditions.

\begin{figure}[!ht]
\centering
\includegraphics[width=1\linewidth]{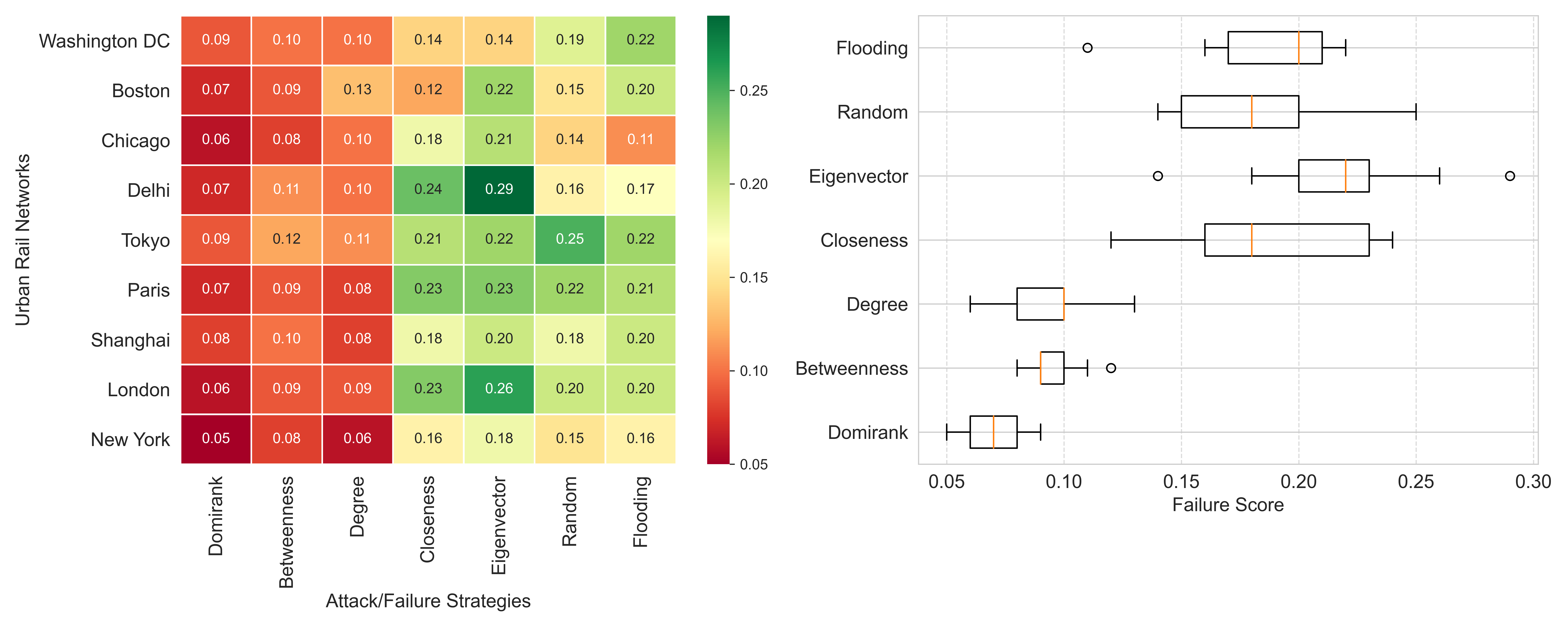}
\caption{\label{fig:Urban}The heatmap on the left displays failure impact scores for nine urban rail networks subjected to disruption strategies, including Domirank, Betweenness, Degree, Closeness, Eigenvector, Random, and Flooding. Lower scores signify a more effective strategy from the attacker's standpoint, implying greater network vulnerability, whereas higher scores reflect a network's robustness. The boxplot on the right details the distribution of these scores, showcasing the variability and relative effectiveness of each strategy. This comparative analysis underscores the diverse resilience profiles across networks, revealing certain networks' susceptibilities, particularly to Domirank and Betweenness strategies, and others' strengths in resisting random and flooding disruptions.}
\end{figure}


\FloatBarrier

\subsection{Recovery}\label{subsec2}

\begin{figure}[!ht]
\centering
\includegraphics[width=1\linewidth]{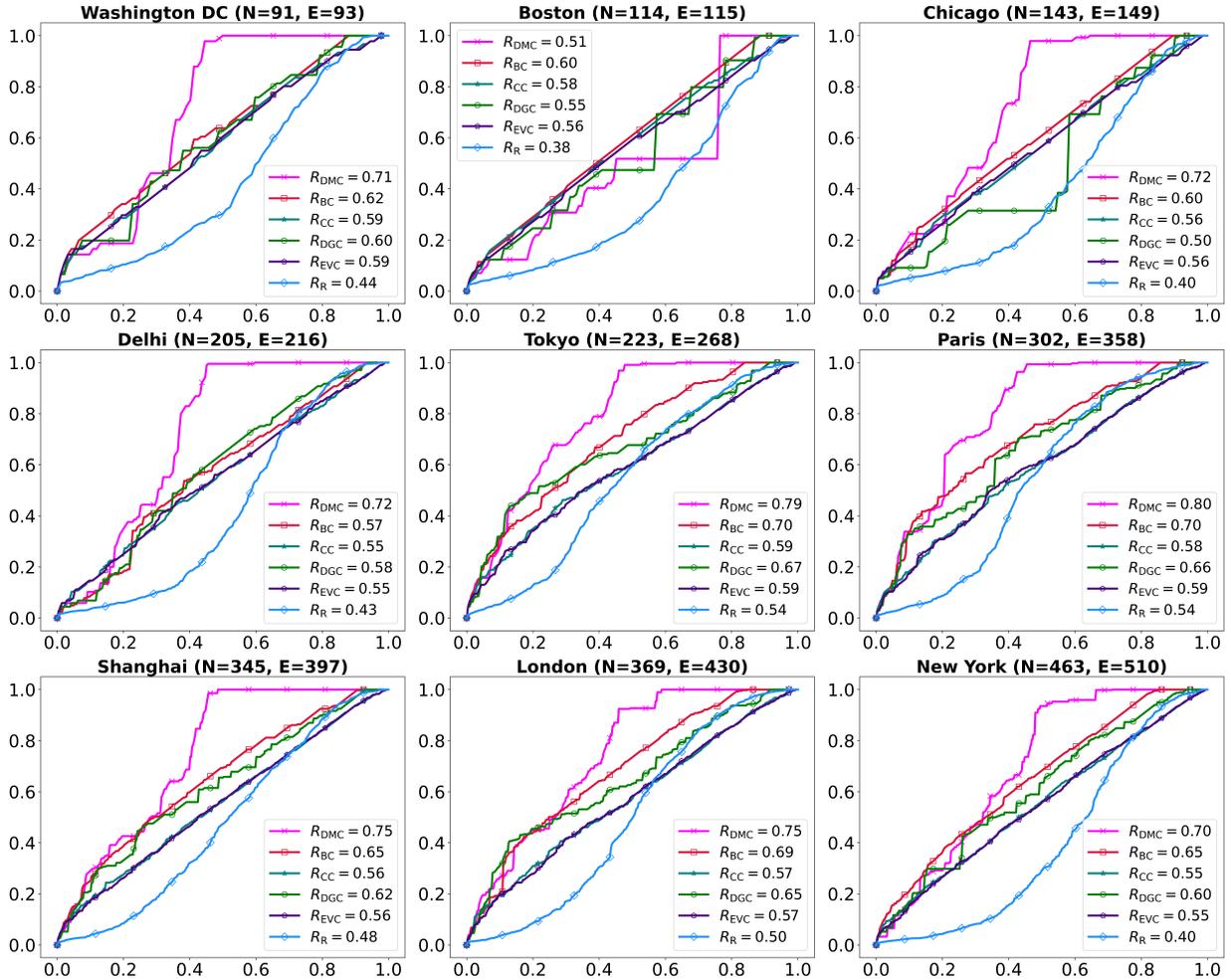}
\caption{\label{fig:Urban}Inter-comparison of recovery strategies of the nine urban rail transportation networks (URTS) after perturbations. The y-axis refers to the relative size of the giant-connected component (GCC) as the network regain functionality and x-axis includes the fraction of nodes (stations) failed till that point. A higher AUC reflects faster recovery of functionality.}
\end{figure}

\begin{figure}[!ht]
\centering
\includegraphics[width=1\linewidth]{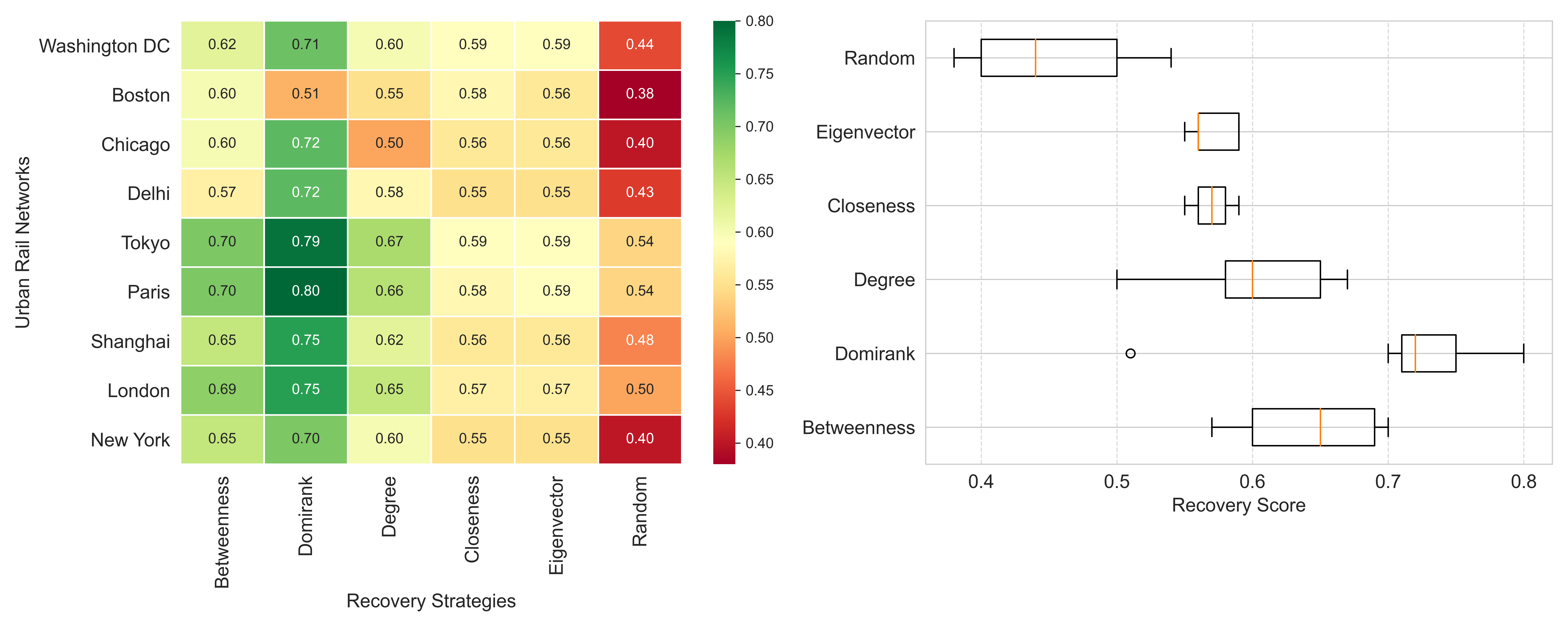}
\caption{\label{fig:Urban}Efficacy of Recovery Strategies in Urban Rail Networks. The heatmap on the left exhibits the recovery scores for nine urban rail networks subjected to various recovery strategies, including Domirank, Betweenness, Degree, Closeness, Eigenvector, and Random. Higher scores indicate a more rapid recovery. The boxplot on the right shows the distribution of recovery scores, with higher median values suggesting more effective recovery. The visualization highlights the differential ability of networks to rebound from disruptions, with certain strategies like Domirank and Betweenness facilitating faster recovery across the analyzed networks.
}
\end{figure}



\begin{figure}[!ht]
\centering
\includegraphics[width=1\linewidth]{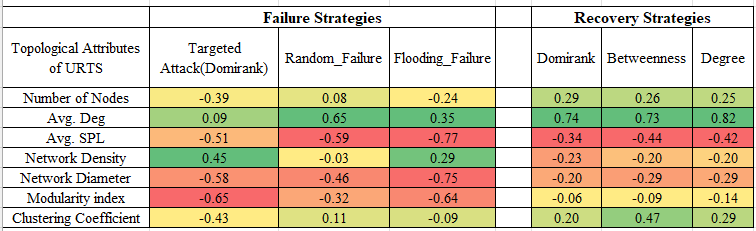}
\caption{\label{fig:Urban}The table presents the Kendall tau correlation coefficients between various network topological metrics and the quantitative scores of disruption and recovery strategies in urban rail networks. The upper section details the impact of Targeted Attack (Domirank), Random Failure, and Flooding Failure on network robustness, while the lower section assesses the efficacy of recovery strategies such as Domirank, Betweenness, and Degree centrality (top three among all strategies). Positive values indicate a direct relationship, whereas negative values suggest an inverse relationship. The color gradient from green to red highlights the strength and direction of each correlation, providing insights into how network characteristics influence the resilience and recovery of urban rail systems.}
\end{figure}

\begin{figure}[!ht]
\centering
\includegraphics[width=1\linewidth]{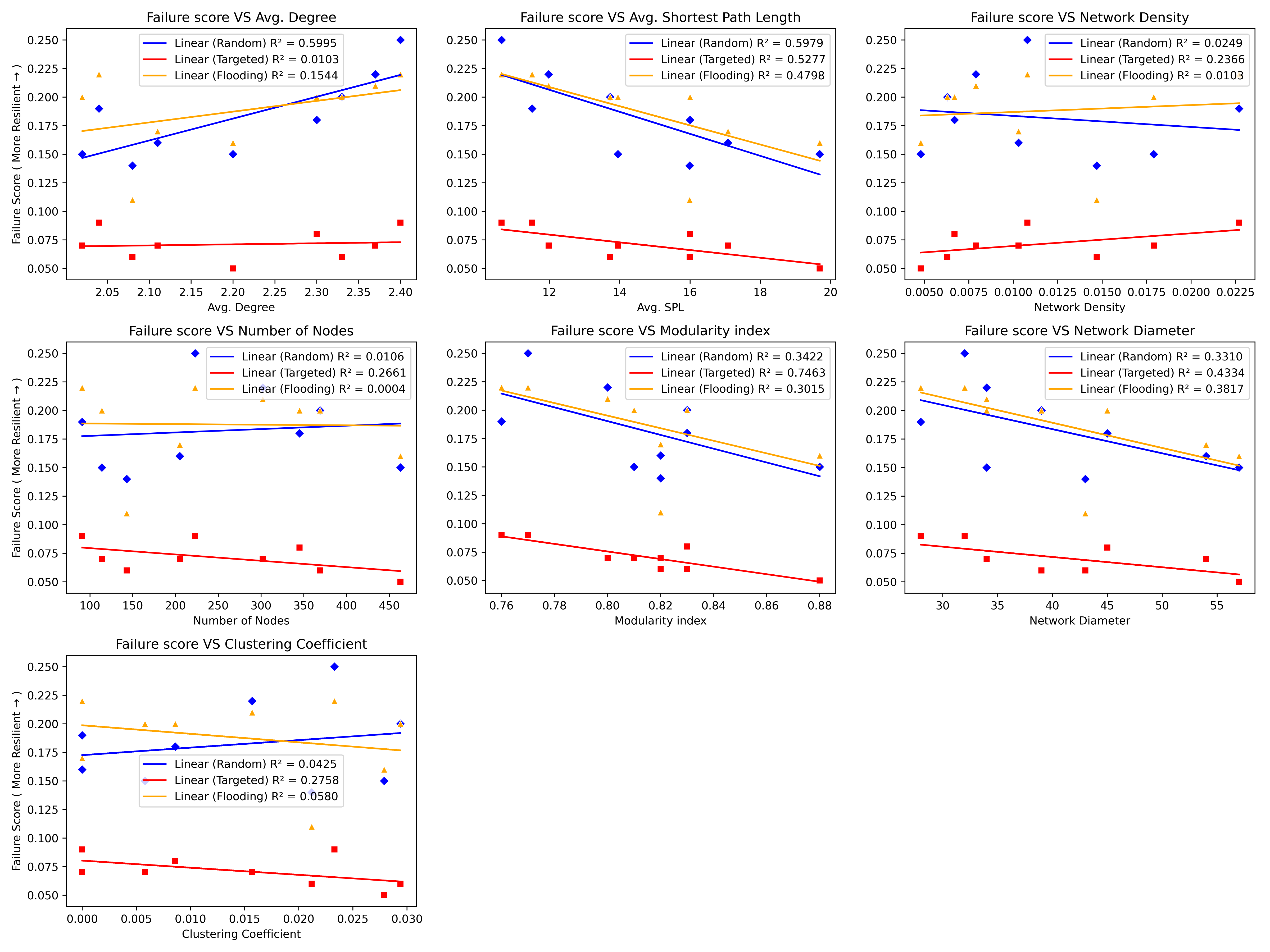}
\caption{\label{fig:Urban}Impact of Network Topology on Urban Rail Systems' Robustness Against Disruptions. These subplots illustrate the correlation between urban rail networks' robustness and key topological metrics: average degree (left) and average shortest path length (right). Data points represent robustness scores under different disruption strategies—Targeted Attack (Domirank), Random Failure, and Flooding Failure—against each network's average degree and shortest path length, with polynomial and linear fit lines indicating trends. The coefficient of determination (R²) for each fit describes the proportion of variance in the robustness score accounted for by the topological metric. A higher R² value suggests a stronger relationship between network topologies.}
\end{figure}

\begin{figure}[!ht]
\centering
\includegraphics[width=1\linewidth]{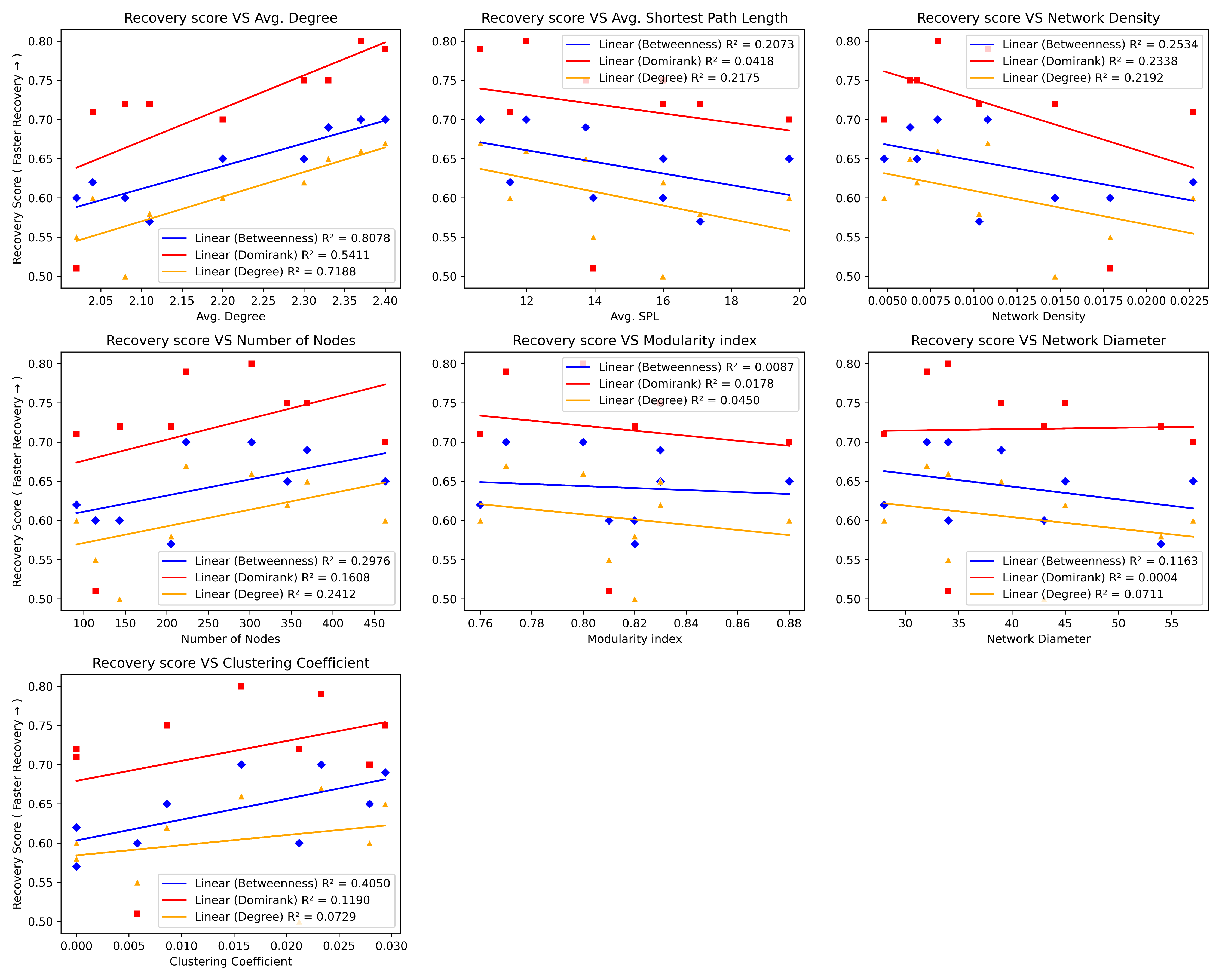}
\caption{\label{fig:Urban}Impact of Network Topology on Urban Rail Systems' recovery strategies from Disruptions. These subplots illustrate the correlation between urban rail networks' robustness and key topological metrics: average degree (left) and average shortest path length (right). Data points represent robustness scores under different disruption strategies—Targeted Attack (Domirank), Random Failure, and Flooding Failure—against each network's average degree and shortest path length, with polynomial and linear fit lines indicating trends. The coefficient of determination (R²) for each fit describes the proportion of variance in the robustness score accounted for by the topological metric. A higher R² value suggests a stronger relationship between network topologies.}
\end{figure}

\FloatBarrier

\section{Discussion:}\label{sec5}

The relationship between network size and network robustness during failures and recovery is multifaceted. On one hand, larger networks are more susceptible to specific types of failures, such as targeted attacks and flooding failures. On the other hand, during recovery, these larger networks show an advantage, as the redundancy and alternative pathways inherent in their structure contribute to greater resilience and an improved ability to recover.
When considering other network properties, certain attributes like average degree and average shortest path length display a consistent influence on network behavior in both failure and recovery scenarios. A higher average degree and a shorter average path length correlate with a network's effectiveness in recovery, highlighting the importance of connectivity and the minimization of node-to-node distances for enhancing network resilience. In contrast, the impact of network density, network diameter, and modularity index on robustness varies with the context, indicating a more complex relationship with network failures.
The clustering coefficient's positive correlation with recovery suggests that networks with greater local interconnectivity among nodes are more robust post-failure. These findings suggest that network designers should carefully balance the design elements, taking into account the trade-offs between the risks posed by larger network sizes in failure scenarios against their recovery capabilities, as well as the impact of other network properties on overall network resilience.

\section{Conclusions}
This study provides a comprehensive examination of the resilience of nine major urban rail networks—Washington DC, Boston, Chicago, Delhi, Tokyo, Paris, Shanghai, London, and New York—against multi-hazard disruptions. Utilizing a quantitative approach focused on topological parameters, we evaluated the failure and recovery phases of these networks through percolation-based network dismantling methods such as Sequential Removal of Nodes and Giant Connected Component analysis.

Our findings indicate that larger urban rail networks, while initially more vulnerable to targeted disruptions and flooding, demonstrate enhanced recovery capabilities due to their inherent redundancy and connectivity. The analysis reveals that key topological attributes such as average degree and path length consistently influence recovery effectiveness. Higher average degrees and shorter path lengths correlate with more effective recovery, highlighting the significance of connectivity and the minimization of node-to-node distances.

The study identifies Domirank centrality as a particularly effective measure for evaluating node importance during both disruption and recovery phases. Networks exposed to Domirank-centrality-based attacks exhibited the lowest robustness, while those subjected to eigenvector-centrality-based disruptions showed greater resilience. The positive correlation of the clustering coefficient with recovery underscores the benefits of local interconnectivity within the network, which enhances resilience by enabling alternative pathways for movement even in the event of partial network failures.

Flooding failures were found to behave similarly to localized random failures but with potentially greater impacts depending on the number of critical nodes in flood-prone areas. The Chicago L network exhibited the lowest resilience to flooding among the nine networks studied, while the Tokyo metro system demonstrated the highest resilience. A consistent pattern emerged across all networks, where a small fraction of nodes played a pivotal role in maintaining network integrity under various types of attacks and failures.

One practical approach to designing a network with higher average degrees and shorter path lengths is the development of a multimodal transit network system. By integrating multiple transit modes such as buses, trains, and subways, the network can maintain functionality even if a single component fails. This multimodal integration enhances overall network resilience by providing alternative routes and connections. Adding another mode of transit to an existing node increases the node's degree and reduces the average shortest path length, thereby improving the network's robustness and efficiency. This analysis of the resilience of multimodal networks could serve as a valuable direction for future research.

\section*{Acknowledgments}
This project was partially funded by the U.S. Department of Defense Strategic Environmental Research and Development
Program project number RC20-1183 titled Networked Infrastructures under Compound Extremes. The views and conclusions in this document are those of the authors and should not be interpreted as necessarily
representing the official policies, either expressed or implied, of the U.S. Department of Defense.

\bibliographystyle{unsrt}  
\bibliography{references}

\end{document}